# Design and Scalable Synthesis of Thermochromic VO$_2$-Based Coatings for Energy-Saving Smart Windows with Exceptional Optical Performance

Michal Kaufman, Jaroslav Vlček,∗ Jiří Houška, Sadoon Farrukh, and Stanislav Haviar
Department of Physics and NTIS-European Centre of Excellence, University of West Bohemia, Univerzitní 8, 30100 Plzeň, Czech Republic
∗Email: vlcek@kfy.zcu.cz

**KEYWORDS:** *doped vanadium dioxide, strongly thermochromic coatings, low transition temperature, scalable synthesis, low deposition temperature, smart windows*

**ABSTRACT:** We report strongly thermochromic YSZ/V$_{0.855}$W$_{0.018}$Sr$_{0.127}$O$_2$/SiO$_2$ coatings, where YSZ is Y stabilized ZrO$_2$, prepared using a scalable deposition technique on standard glass at a low substrate temperature of 320 °C and without any substrate bias voltage. The coatings exhibit a transition temperature of 22 °C with an integral luminous transmittance of 63.7% (low-temperature state) and 60.7% (high-temperature state), and a modulation of the solar energy transmittance of 11.2%. Such a combination of properties, together with the low deposition temperature, fulfill the requirements for large-scale implementation on building glass and have not been reported yet. Reactive high-power impulse magnetron sputtering with a pulsed O$_2$ flow feedback control allows us to prepare crystalline W and Sr co-doped VO$_2$ of the correct stoichiometry. The W doping of VO$_2$ decreases the transition temperature, while the Sr doping of VO$_2$ increases the luminous transmittance significantly. A coating design utilizing a second-order interference in two antireflection layers is used to maximize both the integral luminous transmittance and the modulation of the solar energy transmittance. A compact crystalline structure of the bottom YSZ antireflection layer further improves the VO$_2$ crystallinity, while the top SiO$_2$ antireflection layer provides also the mechanical and environmental protection for the V$_{0.855}$W$_{0.018}$Sr$_{0.127}$O$_2$ layer.

## 1. INTRODUCTION

Global warming and energy crisis drive a focus on energy-saving materials. Buildings have been estimated to produce about 20% of all anthropogenic greenhouse gas emissions[1] and are responsible for up to 40% of the primary energy consumption[2] in the world. Approximately 50% of the total building energy is consumed for compensating the heat gains and losses via windows and glass facades,[3] which are the most energy-inefficient components of buildings. It is evident that energy-saving smart windows with adjustable throughput of solar energy can lower the energy expenditure.[4]

Vanadium dioxide (VO$_2$) exhibits a reversible phase transition from a low-temperature monoclinic VO$_2$ (M1) semiconducting phase to a high-temperature tetragonal VO$_2$ (R) metallic phase at a transition temperature ($T_{tr}$) of approximately 68 °C for the bulk material.[5] The $T_{tr}$ can be lowered using doping of VO$_2$ with other elements (such as W).[6,7] The automatic (i.e., without any switch system) response to temperature and the abrupt decrease of infrared transmittance with almost the same luminous transmittance (allowing to utilize daylight) at the transition into the metallic state make VO$_2$-based coatings a promising candidate for thermochromic smart windows reducing the energy consumption of buildings.

Magnetron sputter deposition with its versatility and ease of scaling up to large substrate sizes is probably the most important preparation technique of thermochromic $VO_2$-based coatings.[7-11] Note that magnetron sputter sources are used very frequently not only in glass production lines (e.g., for deposition of low-emissivity coatings) but also in large-scale roll-to-roll deposition devices[12] producing coatings on ultrathin flexible glass or polymer foils. Moreover, the atom-by-atom magnetron co-sputtering is a much simpler and much more effective method than, e.g., chemical methods, for a doping of the $VO_2$ layers with other elements.

To meet the requirements for large-scale implementation on building glass (glass panes, or flexible glass and polymer foils laminated to glass panes), $VO_2$-based coatings should satisfy the following strict criteria simultaneously: a maximum substrate temperature ($T_s$) during the preparation (deposition and possible post annealing) close to 300 °C or lower,[7-9,13,14] $T_{tr}$ close to 25 °C or lower,[15] an integral luminous transmittance $T_{lum} > 60\%$,[16-18] a modulation of the solar energy transmittance $\Delta T_{sol} > 10\%$,[19-21] long-term environmental stability,[10,22-24] and a more appealing color[25,26] than the usual yellowish or brownish colors in transmission. The simultaneous fulfillment of these requirements has not yet been reported in the literature. A major challenge is to achieve the high $T_{lum}$ and $\Delta T_{sol}$ at a relatively low $T_{tr}$ and $T_s$.[7-27]

Reactive high-power impulse magnetron sputtering (HiPIMS) has been found to be a promising deposition technique for a low-temperature (300 - 350 °C) preparation of undoped thermochromic $VO_2$ films.[13,14,18,22,28,29] In our recent paper,[30] we presented a scalable[31] deposition technique used for a low-temperature preparation of high-performance three-layer $ZrO_2/V_{0.982}W_{0.018}O_2/ZrO_2$ coatings on soda-lime glass (SLG). The thermochromic $V_{0.982}W_{0.018}O_2$ layers were deposited by a controlled HiPIMS of a V target, combined with a simultaneous pulsed DC magnetron sputtering of a W target (doping of $VO_2$ with W to reduce the $T_{tr}$ to 20 °C without any degradation of thermochromic properties), at $T_s$ = 330 °C in an argon-oxygen gas mixture. The coatings exhibited $T_{lum}$ = 49.9% (below the $T_{tr}$) and 46.0% (above the $T_{tr}$), and $\Delta T_{sol}$ = 10.4% for a $V_{0.982}W_{0.018}O_2$ layer thickness of 69 nm.

In this study, we report the design and scalable synthesis of strongly thermochromic $YSZ/V_{0.855}W_{0.018}Sr_{0.127}O_2/SiO_2$ coatings, where YSZ is Y-stabilized $ZrO_2$, which fulfill the aforementioned requirements for large-scale implementation on building glass. The coatings exhibit a transition temperature $T_{tr}$ = 22 °C with an integral luminous transmittance $T_{lum}$ = 63.7% (below the $T_{tr}$) and 60.7% (above the $T_{tr}$), and a modulation of the solar energy transmittance $\Delta T_{sol}$ = 11.2% for a $V_{0.855}W_{0.018}Sr_{0.127}O_2$ layer thickness of 71 nm. We have modified the sputter deposition technique, based on reactive HiPIMS, to perform a controlled co-doping of $VO_2$ with W, shifting the $T_{tr}$ to a room temperature, and with Sr, increasing the $T_{lum}$ substantially. Reactive HiPIMS with a pulsed $O_2$ flow feedback control allowed us to prepare crystalline $VO_2$ of the correct stoichiometry at a low substrate temperature $T_s$ = 320 °C and without any substrate bias voltage. An original design of a three-layer $VO_2$-based coating utilizing a second-order interference in two antireflection (AR) layers was applied to increase both the $T_{lum}$ and the $\Delta T_{sol}$. A compact crystalline structure of the bottom YSZ AR layer further improves the $VO_2$ crystallinity and the process reproducibility, while the top $SiO_2$ AR layer provides also mechanical and environmental protection for the thermochromic $V_{0.855}W_{0.018}Sr_{0.127}O_2$ layer.

## 2. RESULTS AND DISCUSSION

The results presented in this section are for two thermochromic (TC) layers: an optimized $V_{0.855}W_{0.018}Sr_{0.127}O_2$ (see Section 4.1) and $V_{0.984}W_{0.016}O_2$ (prepared for comparative purposes with about the same layer thickness and W content but without Sr). The section is organized as follows. First, we analyze how the Sr incorporation affects the electronic structure and in turn optical constants (Figure 1). Second, we discuss how to translate the benefits of Sr incorporation into as high application potential as possible (Figure 2, Table 1). Third, we present the excellent performance of the subsequently prepared optimized coating and compare it with the state of the art (Figure 3, Table 2).

### 2.1. Effect of W and Sr co-doping on optical band gap and optical properties of two-layer YSZ/$V_{0.855}W_{0.018}Sr_{0.127}O_2$ coating.

The electronic structure of the low-temperature

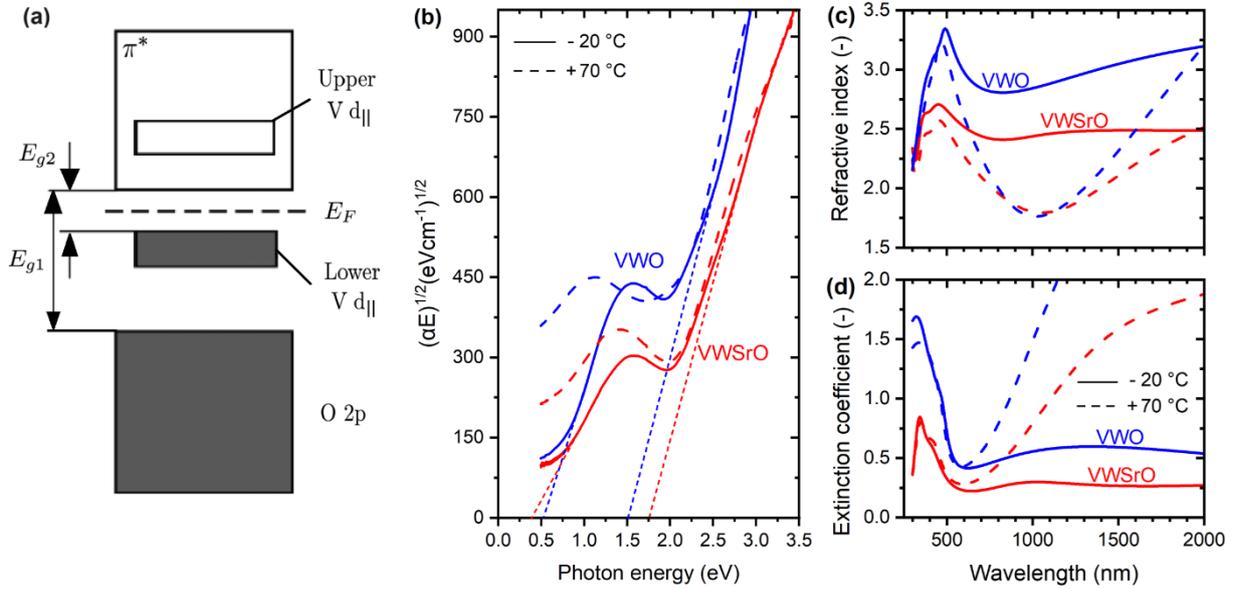

**Figure 1.** a) Schematic energy band diagram with two gaps $E_{g1}$ and $E_{g2}$ for pure $VO_2$(M1). Reproduced with permission.[33] Copyright 2012, AIP Publishing. b) $(\alpha E)^{1/2}$ as a function of the photon energy $E$, where $\alpha$ is the absorption coefficient, for the YSZ(167 nm)/$V_{0.855}W_{0.018}Sr_{0.127}O_2$(71 nm) coating (denoted as VWSrO) and the YSZ (178 nm)/$V_{0.984}W_{0.016}O_2$ (73 nm) coating (denoted as VWO) on 1 mm thick glass at $T_{ms}$ = -20 °C and $T_{mm}$ = 70 °C. At -20 °C, linear fittings are performed to extract $E_{g1}$ and $E_{g2}$ as discussed in the text. Spectral dependences of the refractive index (c) and the extinction coefficient (d) measured for the same two-layer coatings at the same temperatures as in b).

$VO_2$(M1) semiconducting phase is complex, and includes two gaps which affect its functional properties.[32,33] First, there is a band gap in the narrow sense of the word ($E_{g2}$ in Figure 1a) between two bands made up predominantly of V 3d orbitals: the filled lower part of the split $d_{\parallel}$ band and the empty $\pi^*$ band. The width of this gap is in the infrared spectral range with an often reported value of ≈0.6 eV.[34-36] The V atoms are paired and in turn the $d_{\parallel}$ band is split into two only in $VO_2$(M1), while at the transition to $VO_2$(R) this gap closes.[34-36] The consequently enhanced concentration of free charge carriers has a direct effect especially on the contribution of infrared wavelengths to $\Delta T_{sol}$. Second, there is a gap ($E_{g1}$ in Figure 1a) between the second highest filled band made up predominantly of O 2p orbitals and once again the empty $\pi^*$ band. The width of this gap is in the visible range. Therefore, it gives rise to the interband transitions which have a direct effect on $T_{\text{lum}}$ and the coating color, and there are worldwide efforts to improve these characteristics via controlling $E_{g1}$.

The values of $E_{g1}$ and $E_{g2}$, both before ($V_{0.984}W_{0.016}O_2$) and after ($V_{0.855}W_{0.018}Sr_{0.127}O_2$) Sr incorporation, are shown in Figure 1b: Tauc plot $(\alpha E)^{1/2} \sim E - E_g$, where $\alpha$ is the absorption coefficient (neglecting the absorption in glass and YSZ) and the exponent 1/2 is valid for indirect[33,37] allowed transitions. The results of optical measurements are shown for both phases, and the optical gaps of phase M1 are obtained using tangents to linear parts of the low-temperature dependencies. Indeed, the incorporation of 12.7 at.% Sr into the metal sublattice at an almost fixed W content of 1.6-1.8 at.% in the metal sublattice led to a clear widening of the visible-range gap, increasing $E_{g1}$ from 1.51 to 1.75 eV. The trend is in agreement with the previously reported results of doping $VO_2$ with $Sr^{37}$ or co-doping $VO_2$ with W and $Sr^{20}$. A case can be made that the aforementioned elemental compositions are averaged over crystals grains and their amorphous boundaries, i.e., that the true value of the gradient $(1.75-1.51)/12.7 = 0.019$ eV/at.% Sr may be even higher. The enhancement of $E_{g1}$ is also consistent with lowering of the extinction coefficient in the whole visible range, qualitatively observable over here (at a given energy, $\alpha E$ is proportional to $k$) and quantified next. While the figure also indicates that the effect of Sr on $E_{g2}$ is opposite to that on $E_{g1}$, the effect of the infrared-range gap on visible-range properties is arguably less direct, and its narrowing does not constitute a problem as long as the coating performance (Figure 3) is high.

The effect of incorporation of 12.7 at.% Sr into the metal sublattice on $n(\lambda)$ and $k(\lambda)$, as measured by spectroscopic ellipsometry, is shown in Figures 1c and 1d, respectively. On the one hand, both optical constants of both phases exhibit qualitatively similar dispersions with and without Sr. On the other hand, there are important quantitative differences. First, the Sr incorporation leads in most of the wavelength range studied to a lower $n(\lambda)$ of both phases. To put a quantitative example, $n_{550}$ decreased from 2.84 (R) - 3.13 (M1) without Sr to 2.38 (R) - 2.58 (M1) with Sr. This affects not only the best achievable coating performance, but also the way this best performance is achieved (choice of materials for antireflection layers) as discussed in the next section. The observation is consistent with the usual trend: lower polarizability per unit volume and in turn lower refractive index of oxides of main group elements ($n_{550} = 1.88$ for pure SrO)[38] compared to oxides of transition metals. Second and probably most importantly, the Sr incorporation leads to a significantly lower $k(\lambda)$ in the visible. To put a quantitative example, $k_{550}$ decreased from 0.45 (R) - 0.47 (M1) without Sr to 0.29 (R) - 0.26 (M1) with Sr. This confirms the success of the efforts to enhance $E_{g1}$ and opens a pathway toward TC layers possessing higher $T_{\text{lum}}$ (at a given thickness) or higher $\Delta T_{\text{sol}}$ (because lower $k_{550}$ allows a higher thickness) or both. The fact that the lowering of $k_{550}$ of the low-temperature phase M1 is even larger than that of the high-temperature phase R is also beneficial because it can improve the contribution of visible wavelengths to $\Delta T_{\text{sol}}$. Third, the Sr incorporation leads to less steeply increasing $k(\lambda)$ of the metallic phase R in the infrared, i.e., to somewhat weaker TC transition compared to that exhibited by pure or only W-doped $VO_2$. Again, this does not constitute a problem as long as it is more than compensated by the aforementioned benefits and the coating performance (Figure 3) is high.

## 2.2. Design of three-layer YSZ/V$_{0.855}$W$_{0.018}$Sr$_{0.127}$O$_2$/SiO$_2$ coatings.
The design of our TC

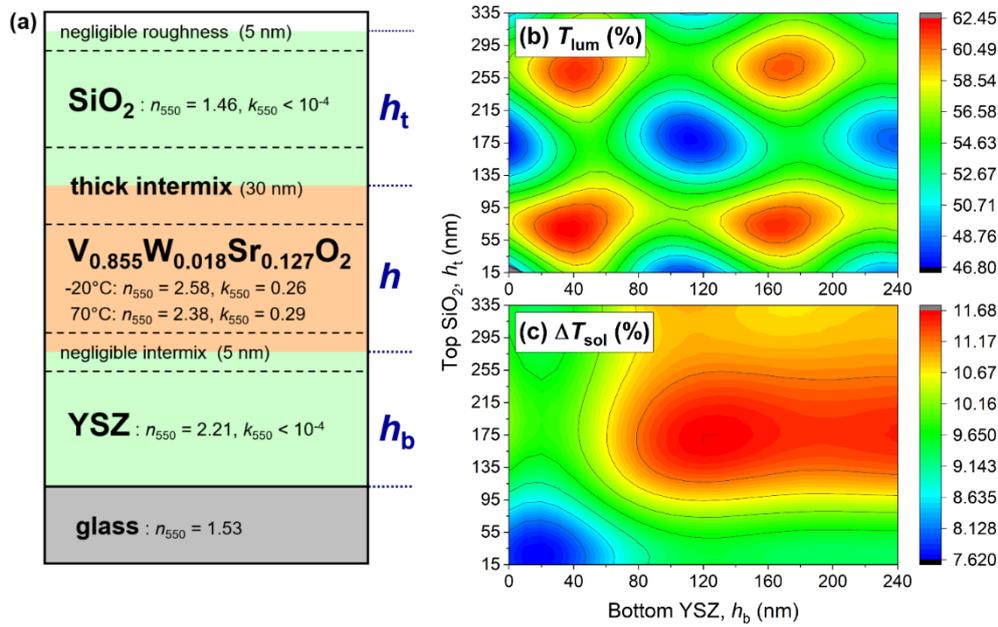

**Figure 2.** a) Optical model used to measure the characteristics of the coating without the top layer (glass/YSZ/V$_{0.855}$W$_{0.018}$Sr$_{0.127}$O$_2$, including the thick surface roughness layer on the V$_{0.855}$W$_{0.018}$Sr$_{0.127}$O$_2$ layer) and to subsequently design the top layer (glass/YSZ/V$_{0.855}$W$_{0.018}$Sr$_{0.127}$O$_2$/SiO$_2$, including the thick intermix layer - instead of the aforementioned surface roughness layer - between the V$_{0.855}$W$_{0.018}$Sr$_{0.127}$O$_2$ and SiO$_2$). b,c) Predicted $T_{lum}$ (average of the similar $T_{lum}$ values at $T_{ms}$ = -20 °C and $T_{mm}$ = 70 °C) and $\Delta T_{sol}$, respectively, as a function of the thickness of the bottom YSZ ($h_b$) and the top SiO$_2$ ($h_t$) for the V$_{0.855}$W$_{0.018}$Sr$_{0.127}$O$_2$ thickness $h$ = 71 nm (value measured before the deposition of the top layer: 56 nm bulk + half of the 30 nm roughness).

coating is shown in Figure 2a. The overall idea has been utilized[30,31,39,40] previously: the active TC layer (thickness $h$) is combined with a bottom AR layer (which constitutes also a crystalline template; thickness $h_b$) and a top AR layer (which constitutes also a mechanical and environmental protection; thickness $h_t$). In order to achieve as high efficiency of the AR layers (as high upper envelope of $T(\lambda)$) as possible, their refractive indices should be between that of glass and TC layer (bottom AR layer) and between that of air and TC layer (top AR layer), ideally (neglecting finite $h$ and $k(\lambda)$ of the TC layer a square root of the corresponding product. The pure ZrO$_2$[30,31,39] or Y-stabilized ZrO$_2$[40], used by us previously as a material for both AR layers around Sr-free W-doped VO$_2$, possesses acceptable $n(\lambda)$ and negligible $k(\lambda)$ in the visible, high hardness (for an oxide), easily achievable crystallinity and controllable crystal orientation, and high potential to serve as a crystalline template (especially in the case of YSZ). In our recent work,[40] we showed that the crystal structure of the bottom YSZ layer, formed by tetragonal YSZ crystal grains, is much more compact (very narrow amorphous boundary regions) than that of the bottom ZrO$_2$ layer comprising monoclinic and tetragonal ZrO$_2$ crystal grains.

However, there are also new issues to consider resulting from the incorporation of Sr. First, the lowered $n_{550}$ of the TC layer (2.38 to 2.58; Figure 1c) leading to a lower optimum $n_{550}$ of the top AR layer ($\sqrt{2.38}$ = 1.54 to $\sqrt{2.58}$ = 1.61) makes high refractive index materials such as YSZ

(in our case $n_{550} = 2.21$) rather suboptimum for the top AR layer, and at the same time increases the comparative advantage of low refractive index materials such as $SiO_2$ ($n_{550} = 1.46$). This has been confirmed by a comparison of results of optical modelling for different designs (not shown). Thus, the optical modelling is shown (Figures 2b,c) and the experiments have been performed (Figure 3) for the optimum $YSZ/V_{0.855}W_{0.018}Sr_{0.127}O_2/SiO_2$ coating. Second, it has been found that the surface roughness of sputtered $VO_2$ co-doped with W and Sr is much higher ($\approx 30$ nm) than the almost negligible roughness of $VO_2$ doped only with W ($\approx 10$ nm) or YSZ and $SiO_2$ ($\approx 5$ nm). This roughness is not negligible anymore, and it has been included in the optical modelling of glass/$YSZ/V_{0.855}W_{0.018}Sr_{0.127}O_2/SiO_2$ by inserting a 30 nm thick intermix layer between $V_{0.855}W_{0.018}Sr_{0.127}O_2$ and $SiO_2$. The properties of this intermix layer are given by 50 vol.% of each of these two materials (Bruggemann effective medium approximation), and the layer contributes by 15 nm to both $h$ and $h_t$.

After choosing candidate materials for AR layers, it is necessary to identify optimum thicknesses of these layers. This has been done by optical modelling using our own code based on a transfer matrix formalism, and the results are shown in Figures 2b,c for the thickness of the TC layer measured before the design and deposition of the top layer: $h = 56$ nm bulk + half of the 30 nm roughness = 71 nm. Figure 2b shows the first-order and second-order interference maxima of $T_{lum}$. While the height of all these maxima is comparable, it is necessary to choose that one which combines high $T_{lum}$ with high $\Delta T_{sol}$. After complementing $T_{lum}$ in Figure 2b with $\Delta T_{sol}$ in Figure 2c, it can be seen that their best combination is associated with two second-order (three-quarter wavelength) AR layers, specifically $h_b = 170$ nm of YSZ and $h_t = 269$ nm of $SiO_2$. The reason[39] is the following: while the first-order maximum in the visible does not lead to anything special in the infrared, the second-order maximum in the visible leads to a first-order maximum in the infrared (at $\approx 3\times$ longer wavelength). The enhanced transmittance in the infrared leads also to enhanced transmittance modulation in the infrared, and in turn to enhanced contribution of infrared wavelengths to $\Delta T_{sol}$. Thus, this is the optimized design used in our experiments.

Furthermore, while the presented recommendation $h_b = 170$ nm of YSZ and $h_t = 269$ nm is valid for $h = 71$ nm, it is worth it to investigate a possible negative correlation between a chosen $h$ and optimum $h_b$ and $h_t$. The results of this investigation are shown in Table 1. On the one hand, because of very different refractive indices of $V_{0.855}W_{0.018}Sr_{0.127}O_2$ and $SiO_2$, the correlation of $h$ and optimum $h_t$ is very weak. On the other hand, because of relatively similar refractive indices of $V_{0.855}W_{0.018}Sr_{0.127}O_2$ and YSZ, the interference in the corresponding bilayer becomes important and leads to a considerable negative correlation of $h$ increasing from 40 to 100 nm and optimum $h_b$ decreasing from 190 to 160 nm. In parallel, Table 1 quantifies the $h$-dependent tradeoff between achievable $T_{lum}$ and $\Delta T_{sol}$: increasing $h$ from 40 to 100 nm combined with optimum $h_b$ and $h_t$ leads to decreasing predicted $T_{lum}$ from 72.6% to 49.6% and to increasing predicted $\Delta T_{sol}$ from 7.41% to 13.61%.

**Table 1.** Optimum thickness of the bottom second-order antireflection YSZ layer, optimum thickness of the top second-order antireflection SiO$_2$ layer, and the $T_{lum}$ and $\Delta T_{sol}$ values which these layers lead to. The data are shown for the thickness of the thermochromic V$_{0.855}$W$_{0.018}$Sr$_{0.127}$O$_2$ layer (denoted as VWSrO) from 40 to 100 nm.

| VWSrO (nm) | YSZ (nm) | SiO$_2$ (nm) | $T_{lum}$ (%) | $\Delta T_{sol}$ (%) |
|---|---|---|---|---|
| 40 | 190 | 274 | 72.6 | 7.41 |
| 55 | 179 | 270 | 66.9 | 9.22 |
| 70 | 171 | 269 | 61.4 | 10.89 |
| 85 | 163 | 273 | 55.5 | 12.38 |
| 100 | 160 | 280 | 49.6 | 13.61 |

**2.3. Structure and thermochromic properties of three-layer YSZ/V$_{0.855}$W$_{0.018}$Sr$_{0.127}$O$_2$/SiO$_2$ coatings.** We have followed the presented design and prepared

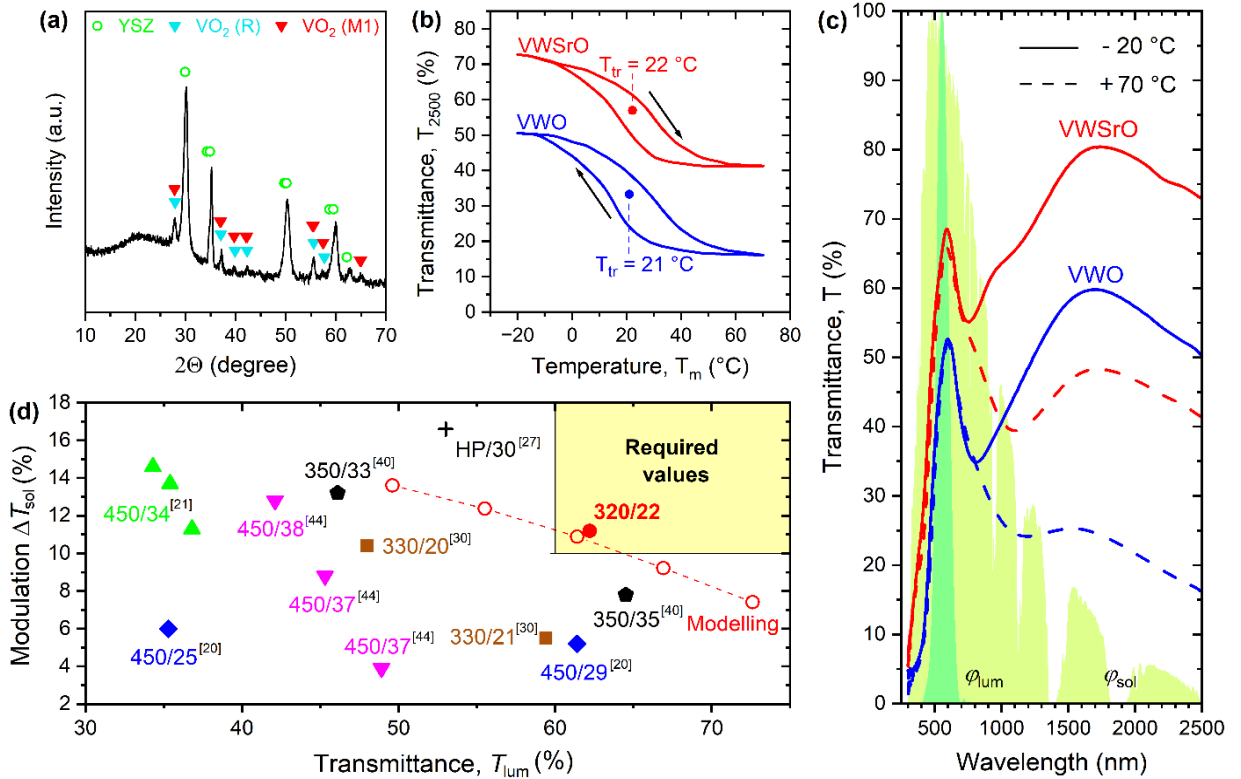

**Figure 3.** a) X-ray diffraction pattern taken at $T_m$ = 25 °C from the YSZ(167 nm)/V$_{0.855}$W$_{0.018}$Sr$_{0.127}$O$_2$(71 nm)/SiO$_2$(280 nm) coating on 1 mm thick glass. The main diffraction peaks of VO$_2$(M1), VO$_2$(R) and YSZ (tetragonal Y$_{0.06}$Zr$_{0.94}$O$_{1.97}$) are marked. b) Temperature dependence of the transmittance at $\lambda$ = 2500 nm for the YSZ(167 nm)/V$_{0.855}$W$_{0.018}$Sr$_{0.127}$O$_2$(71 nm)/SiO$_2$(280 nm) coating (denoted as VWSrO) and the YSZ(178 nm)/V$_{0.984}$W$_{0.016}$O$_2$(73 nm)/SiO$_2$(280 nm) coating (denoted as VWO). The transition temperatures are also given. c) Spectral transmittance measured for the same three-layer coatings as in b) at $T_{ms}$ = -20 °C and $T_{mm}$ = 70 °C. The contours of the shaded areas represent the luminous sensitivity of the human eye ($\varphi_{lum}$) and the solar irradiance spectrum ($\varphi_{sol}$),

normalized to maxima of 100%. d) The average luminous transmittance and the modulation of the solar energy transmittance achieved in this work (full circle) and reported in the literature[20,21,30,40,44] for $VO_2$-based coatings with a transition temperature $T_{tr} \leq 38$ °C prepared on glass substrates using magnetron sputter deposition. Adapted with permission.[40] Copyright 2023, ELSEVIER. The labels denote a maximum substrate temperature during the preparation (deposition and post annealing) of the coatings and their transition temperature (both in °C). For comparison, we give an excellent result (marked with a cross) achieved recently using a hydrothermal process (HP).[27] The shaded area represents the required values of $T_{lum}$ and $\Delta T_{sol}$ for smart-window applications. Calculated $T_{lum}$ and $\Delta T_{sol}$ (empty circles) for the $V_{0.855}W_{0.018}Sr_{0.127}O_2$ thickness from 40 nm (on the right side) to 100 nm with a step of 15 nm are also presented (see Table 1).

TC coatings YSZ(167 nm)/$V_{0.855}W_{0.018}Sr_{0.127}O_2$(71 nm)/$SiO_2$(280 nm) and (for comparative purposes) YSZ(178 nm)/$V_{0.984}W_{0.016}O_2$(73 nm)/$SiO_2$(280 nm) with thicknesses of second-order AR layers very close to the optima given in the previous section. The crystalline phases identified by room-temperature XRD in the former coating can be seen in Figure 3a. The YSZ layer contributes by strong diffraction peaks close to the positions reported for tetragonal $Y_{0.06}Zr_{0.94}O_{1.97}$ (PDF #04-021-9607),[41] confirming its good crystallinity and its ability to serve as a crystalline template.[40] The TC layer contributes by diffraction peaks very close to the positions of both $VO_2$(M1) (PDF #04-003-2035) and $VO_2$(R) (PDF #01-073-2362). These two desired phases are difficult to distinguish, let alone prone to be present simultaneously due to the $T_{tr}$ (see Figure 3b) very close to the XRD measurement temperature. A weak contribution of the amorphous $SiO_2$ layer cannot be distinguished from that of the glass substrate. The most important piece of information in Figure 3a is the absence of any other peaks: there are no fingerprints of non-thermochromic (in the $T_m$ range of interest) stoichiometries such as $V_2O_3$ or $V_4O_9$, or polymorphs such as $VO_2$(P) or $VO_2$(B). This confirms the success of the used pulsed $O_2$ flow feedback process control and of our sputter deposition technique in general.

The TC transition temperature has been examined by measuring the temperature dependence of $T_{2500}$. The hysteresis curves obtained, both with and without Sr, are shown in Figure 3b. It can be seen that the doping with 1.6-1.8 at.% W in the metal sublattice (that is, destabilization of the low-temperature semiconducting phase by the larger size and extra valence electron of W compared to V) allowed us to lower $T_{tr}$ from ≈57 °C (HiPIMS deposition of pure $VO_2$)[28] to desired 21-22 °C. This role of W is not only qualitatively but also almost quantitatively independent of the presence of Sr, leading to a slightly narrower hysteresis curve. Let us emphasize that (contrary to some other results or even generalizing statements in the literature)[42-44] the sputter deposition technique used allowed us to lower $T_{tr}$ at preserved strongly TC behavior.

The TC behavior is quantified in Figure 3c in terms of low- and high-temperature spectral transmittance, once again both with and without Sr. First, the figure captures the role of AR layers: there are second-order maxima of $T(\lambda)$ at ≈600 nm (intentionally not at 550 nm, because the absorption around these maxima is lower at higher $\lambda$) as well as first-order maxima around ≈1700 nm. Second, the figure captures the role of Sr, leading, in agreement with the presented enhancement of $E_{g1}$ in Figure 1b and lowering of $k(\lambda)$ in Figure 1d, to significantly enhanced $T(\lambda)$ (at about the same $h$) at both measurement temperatures. While the transmittance enhancement is arguably most important in the visible, it takes place in the whole $\lambda$ range investigated. Third, while both coatings exhibit $T(\lambda)$ modulation in the infrared, it is welcomed that the Sr incorporation led to a stronger modulation at the shortest infrared wavelengths (where it is multiplied by higher $\varphi_{sol}$) at a cost of weaker modulation at longer wavelengths (where it is multiplied by lower $\varphi_{sol}$).

The spectral transmittance was used to calculate the integral transmittances $T_{lum}$ and $T_{sol}$, and their modulations. The results are provided in Figure 3d and Table 2. The table shows that the Sr incorporation led to a significant enhancement of both key quantities: $T_{lum}$ increased from 45.3% to 60.7% (high-temperature state) or even from 47.0% to 63.7% (low-temperature state), and $\Delta T_{sol}$ increased from 8.4% to 11.2%. The enhancement of $T_{lum}$ is due to the Sr-induced lowering of $k(\lambda)$ in the visible, while the enhancement of $\Delta T_{sol}$ is due to the contribution of both visible wavelengths (because the Sr-induced lowering of $k(\lambda)$ is larger in the low-temperature state, see the enhancement of $\Delta T_{lum}$ from 1.7% to 3.0%) and the shortest infrared wavelengths (because of the Sr-induced stronger modulation of $T(\lambda)$). Thus, the Sr incorporation allowed us to fulfill all quantitative criteria for large-scale implementation of these energy-saving TC coatings (Section 1): not only $T_s$ close to 300 °C and $T_{tr}$ close to 25 °C, but also $T_{lum}$ > 60% and $\Delta T_{sol}$ > 10%. Let us emphasize that the criteria have been fulfilled not only in terms of average $T_{lum}$ (used by most available papers) but also in terms of minimum $T_{lum}$ (which is arguably more relevant). The comparison with the literature in Figure 3d proves that the criteria of success were not only fulfilled but fulfilled for the first time. Note that this has been allowed not only by the proper coating design, deposition technique and elemental composition of the TC layer, but also by the proper thickness of the TC layer. The dashed line in Figure 3d (visualization of the data from Table 1) in a close vicinity of the experimental datapoint not only confirms the correctness of the ellipsometric measurements and optical modelling, but also shows that $h$ = 71 nm is close to the middle of the narrow $h$ range where both $T_{lum}$ and $\Delta T_{sol}$ are sufficiently high. On the contrary, all VO$_2$-based coatings reported in the literature fail to meet the requirement for at least one of the quantities from the quadruplet $T_s$, $T_{tr}$, $T_{lum}$, and $\Delta T_{sol}$ (let alone reports and coating comparisons which do not even mention some of them). A case can be made that there is a coating recently prepared[27] using a rather complicated hydrothermal synthesis which would possibly almost fulfill these conditions (at $T_{tr}$ = 30 °C) in a case of different $h$ choice, albeit the scalability of the process to large deposition devices is (contrary to magnetron sputter deposition of VO$_2$-based coatings)[31] yet to be demonstrated.

**Table 2.** The integral luminous and solar energy transmittance ($T_{lum}(T_m)$ and $T_{sol}(T_m)$, respectively) measured at $T_{ms}$ = -20 °C and $T_{mm}$ = 70 °C, together with the corresponding modulations $\Delta T_{lum}$ and $\Delta T_{sol}$, for the YSZ(167 nm)/V$_{0.855}$W$_{0.018}$Sr$_{0.127}$O$_2$(71 nm)/SiO$_2$(280 nm) coating (denoted as VWSrO) and the YSZ(178 nm)/V$_{0.984}$W$_{0.016}$O$_2$(73 nm)/SiO$_2$(280 nm) coating (denoted as VWO) on 1 mm thick glass.

| Sample | $T_{lum}(T_{ms})$ (%) | $T_{lum}(T_{mm})$ (%) | $\Delta T_{lum}$ (%) | $T_{sol}(T_{ms})$ (%) | $T_{sol}(T_{mm})$ (%) | $\Delta T_{sol}$ (%) |
|---|---|---|---|---|---|---|
| VWSrO | 63.7 | 60.7 | 3.0 | 58.8 | 47.6 | 11.2 |
| VWO | 47.0 | 45.3 | 1.7 | 39.9 | 31.5 | 8.4 |

As expected (see Figure 3c), no visible change in the transparency and color of the strongly thermochromic YSZ/V$_{0.855}$W$_{0.018}$Sr$_{0.127}$O$_2$/SiO$_2$ coating is observed in Figure 4 when its temperature increased from 5 °C to 55 °C.

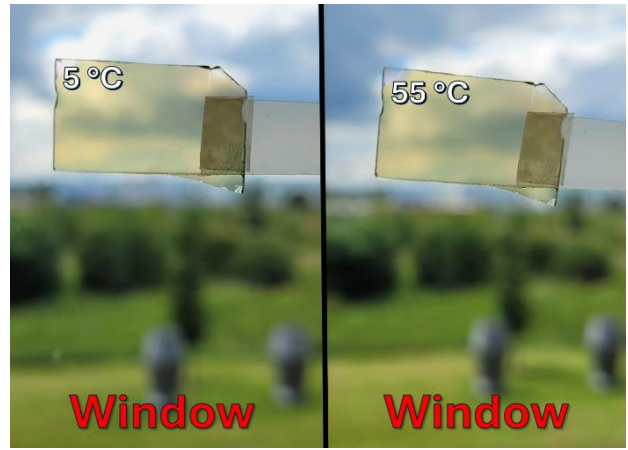

**Figure 4.** View of a YSZ(167 nm)/V$_{0.855}$W$_{0.018}$Sr$_{0.127}$O$_2$(71 nm)/SiO$_2$(280 nm) coating with $T_{tr}$ = 22 °C on 1 mm thick glass at $T_m$ = 5 °C and 55 °C, attached to a window to illustrate its transparency and color.

## 3. CONCLUSIONS

Strongly thermochromic energy-saving YSZ/V$_{0.855}$W$_{0.018}$Sr$_{0.127}$O$_2$/SiO$_2$ coatings have been prepared using a scalable deposition technique on conventional glass. All quantitative criteria for large-scale implementation of these coatings have been fulfilled simultaneously for the first time: $T_s$ = 320 °C without any substrate bias voltage, $T_{tr}$ = 22 °C, $T_{lum}$ = 60.7% (high-temperature state) to 63.7% (low-temperature state) and $\Delta T_{sol}$ = 11.2%. The success has been achieved by a combining: i) full utilization of the advantages of reactive HiPIMS deposition with the effective pulsed O$_2$ flow feedback control, ii) coating design with second-order AR layers, iii) proper choice of the materials for AR layers (YSZ and SiO$_2$), iv) optimum level of doping the metal sublattice of VO$_2$ with W (1.8 at.% in order to destabilize the low-temperature phase and lower $T_{tr}$), and v) optimum level of doping the metal sublattice of VO$_2$ with Sr (12.7 at.% in order to widen the visible-range optical gap). This moves us closer to reducing the energy consumption of buildings by applying this kind of coatings on windows and glass facades.

## 4. EXPERIMENTAL SECTION

**4.1. Coating Preparation.** The coatings were deposited onto 1 mm thick SLG substrates in argon-oxygen gas mixtures at the argon partial pressure $p_{Ar}$ = 1 Pa, corresponding to the argon flow rate of 60 sccm, in an ultra-high vacuum multi-magnetron sputter device (ATC 2200-V AJA International Inc.) equipped by unbalanced magnetrons with planar targets (diameter of 50 mm and thickness of 6 mm in all cases). The base pressure before deposition was below 10$^{-4}$ Pa. The rotating (20 rpm) substrates at the distance of 145 mm from the targets were at a floating potential. The V$_{0.855}$W$_{0.018}$Sr$_{0.127}$O$_2$ layer was deposited by controlled HiPIMS of a single V-W (4.0 wt.% corresponding to 1.14 at.%) target (99.95% purity), combined with a simultaneous pulsed DC magnetron sputtering of a Sr target (99.8% purity), at the substrate surface temperature $T_s$ = 320 °C (Figure 5).

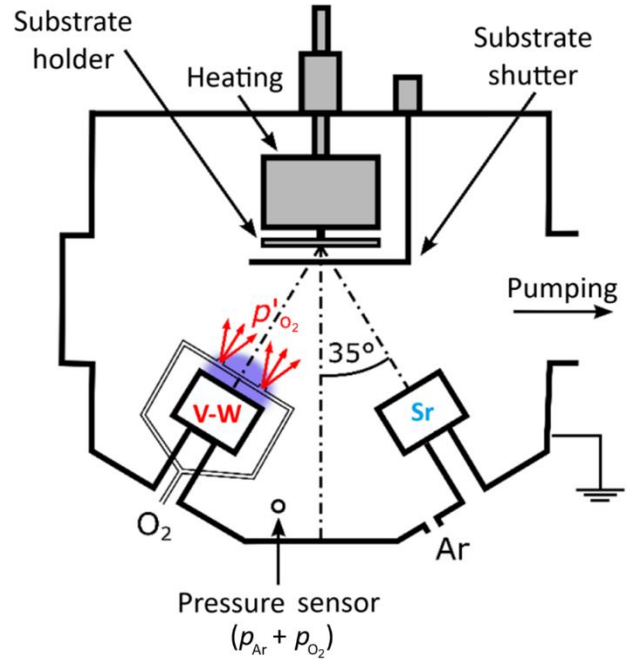

Figure 5. Schematic diagram of the deposition system with four magnetron targets showing two magnetrons with V-W and Sr targets, located in opposite positions, which were used for the deposition of the $V_{0.855}W_{0.018}Sr_{0.127}O_2$ layer. Two $O_2$ inlets were placed 20 mm from the V-W target surface and oriented to the substrate. Positions of the pressure sensor and the Ar inlet at the wall of the vacuum chamber are also shown. An increased local value of the oxygen partial pressure due to the $O_2$ injection is denoted as $p'_{O_2}$.

The total oxygen flow rate ($\Phi_{O_2}$) in two to-substrate $O_2$ inlets, injecting oxygen in front of the V-W target, was not fixed but alternating between 2.0 sccm and 3.0 sccm. This resulted in oscillations of the oxygen partial pressure $p_{O_2}$ between 26 mPa and 55 mPa. The moments of switching of the $\Phi_{O_2}$ pulses were determined during the deposition by a programmable logic controller using a pre-selected critical value of the discharge current on the V-W target. The basic principles of this effective pulsed oxygen flow control are presented in our recent papers.[28,30,40] The magnetron with the V-W target was driven by a unipolar high-power pulsed DC power supply (TruPlasma Highpulse 4002 TRUMPF Huettinger). The voltage pulse duration was 80 μs at a repetition frequency of 500 Hz (duty cycle of 4%) and the deposition-averaged target power density (spatially averaged over the total target area) was 14.9 Wcm$^{-2}$. The magnetron with the Sr target was driven by a unipolar pulsed DC power supply (IAP-1010 EN Technologies Inc.). To minimize arcing on the Sr target surface at an increased Sr target power density and to control a Sr content in the layers easily, we used short 7μs voltage pulses at a relatively high repetition frequency of 50 kHz (duty cycle of 35%) during the depositions with a pre-selected deposition-averaged target power density in the range from 0.1 Wcm$^{-2}$ to 3.3 Wcm$^{-2}$. The optimized thermochromic $V_{0.855}W_{0.018}Sr_{0.127}O_2$ layer with a thickness of 71 nm was deposited onto a 167 nm thick YSZ layer on a 1 mm thick SLG substrate at the deposition-averaged target power density of 1.9 Wcm$^{-2}$. It exhibited (without a top $SiO_2$ AR layer) a high

$T_{lum}$ = 56.8% (low-temperature state) and the highest achieved $\Delta T_{sol}$ = 8.3%. A further increase in the Sr content resulted in a higher $T_{lum}$ and a lower $\Delta T_{sol}$. Details of the developed sputter deposition technique and the effect of an increasing Sr content in the W and Sr co-doped $VO_2$ films on their electronic and crystal structure, and time-dependent optical and electrical properties will be presented elsewhere. The $V_{0.984}W_{0.016}O_2$ layer was deposited by controlled HiPIMS of a V target (99.9% purity), combined with a simultaneous pulsed DC magnetron sputtering of a W target (99.9% purity), at $T_s$ = 320 °C. The $\Phi_{O_2}$ and $p_{O_2}$ oscillated between 1.5 sccm and 1.9 sccm, and between 23 mPa and 72 mPa, respectively, during the deposition performed using the same power supplies. For the V target (HiPIMS), the voltage pulse duration was 80 µs at a repetition frequency of 500 Hz and the deposition-averaged target power density was 14.2 Wcm$^{-2}$. For the W target, the voltage pulse duration was 16 µs at a repetition frequency of 5 kHz and the deposition-averaged target power density was 25 mWcm$^{-2}$. The YSZ layers were deposited by controlled HiPIMS of a single Zr-Y (9.0 wt.% corresponding to 9.2 at.%) target (99.9% purity) at $T_s$ = 320 °C. The $\Phi_{O_2}$ and $p_{O_2}$ oscillated between 1.9 sccm and 2.4 sccm, and 25 mPa and 73 mPa, respectively, during the depositions performed using the aforementioned high-power pulsed DC power supply. The voltage pulse duration was 80 µs at a repetition frequency of 500 Hz and the deposition-averaged target power density was 14.9 Wcm$^{-2}$. The $SiO_2$ layers were deposited by mid-frequency bipolar dual magnetron sputtering of two Si (99.999% purity) targets at $T_s \leq 35$ °C (without any external heating). The $\Phi_{O_2}$ = 17 sccm and $p_{O_2}$ = 0.2 Pa during the depositions performed using a bipolar dual power supply (TruPlasma Bipolar 4010 TRUMPF Huettinger). The voltage pulse duration was 10 µs at a repetition frequency of 50 kHz and the deposition-averaged target power density was approximately 8 Wcm$^{-2}$.

**4.2. Coating Characterization.** The W and Sr contents in the metal sublattice of $V_{0.984}W_{0.016}O_2$ and $V_{0.855}W_{0.018}Sr_{0.127}O_2$, i.e., 1.6 ± 0.3 at.% of W, and 1.8 ± 0.2 at.% of W and 12.7 ± 1.8 at.% of Sr, respectively, were measured on a dedicated 440 nm thick layer on Si(100) substrate in a scanning electron microscope (SU-70, Hitachi) using wave-dispersive spectroscopy (Magnaray, Thermo Scientific) at a low primary electron energy of 7.5 keV. Standard reference samples of pure V, W, $Fe_2O_3$ and $SrSO_4$ (Astimex Scientific Ltd.) were utilized. The room-temperature (25 °C) crystal structure of coatings was characterized by X-ray diffraction using a PANalytical X'Pert PRO diffractometer working with a CuKα (40 kV, 40 mA) radiation at a glancing incidence of 1°. The thickness and optical constants (refractive index, $n$, and extinction coefficient, $k$) of individual layers were measured by spectroscopic ellipsometry using the J.A. Woollam Co. Inc. VASE instrument equipped by an Instec heat/cool stage. The measurements of $n(\lambda)$ and $k(\lambda)$ were performed in the wavelength range 300 – 2000 nm at the angles of incidence of 55°, 60° and 65° in reflection for $T_{ms}$ = -20 °C (semiconducting state below $T_{tr}$) and $T_{mm}$ = 70 °C (metallic state above $T_{tr}$). YSZ was described by the Cauchy dispersion formula, and $V_{0.984}W_{0.016}O_2$ and $V_{0.855}W_{0.018}Sr_{0.127}O_2$ were represented by a combination of the Cody-Lorentz oscillator, Lorentz oscillators and (in the case of metallic phase) Drude oscillator. The coating transmittance ($T$) and reflectance ($R$) were measured by spectrophotometry using the Agilent CARY 7000 instrument with an in-house made heat/cool cell. The measurements were performed in the wavelength range 300 – 2500 nm at the angles of incidence of 0° ($T$) and 7° ($R$) for $T_{ms}$ = -20 °C and $T_{mm}$ = 70 °C. Hysteresis curves were measured for $T$ at $\lambda$ = 2500 nm in the temperature range $T_m$ = -20 °C to 70 °C. The coating performance is quantified by means of integral luminous transmittance ($T_{lum}$), integral solar energy transmittance ($T_{sol}$) and their modulations ($\Delta T_{lum}$ and $\Delta T_{sol}$). The quantities are defined as

$$T_{lum}(T_m) = \frac{\int_{380}^{780} \varphi_{lum}(\lambda)\varphi_{sol}(\lambda)T(T_m,\lambda)d\lambda}{\int_{380}^{780} \varphi_{lum}(\lambda)\varphi_{sol}(\lambda)d\lambda},$$

$$\Delta T_{lum} = T_{lum}(T_{ms}) - T_{lum}(T_{mm}),$$

$$T_{sol}(T_m) = \frac{\int_{300}^{2500} \varphi_{sol}(\lambda)T(T_m,\lambda)d\lambda}{\int_{300}^{2500} \varphi_{sol}(\lambda)d\lambda},$$

$$\Delta T_{sol} = T_{sol}(T_{ms}) - T_{sol}(T_{mm}),$$

where $\varphi_{lum}$ is the luminous sensitivity of human eye and $\varphi_{sol}$ is the solar irradiance spectrum at an air mass of 1.5.[45] The average luminous transmittance is defined as $T_{lum} = [T_{lum}(T_{ms}) + T_{lum}(T_{mm})]/2$. The optical band gaps $E_{g1}$ and $E_{g2}$ were determined from Tauc plots by utilizing the relation $(\alpha E)^{1/2} \sim E - E_g$,[33,37] where $E$ is the photon energy and $\alpha$ is the absorption coefficient calculated as $\alpha = -[\ln(T/1-R)]/h$,[33] where $h$ is the thickness of TC layer.

## AUTHOR INFORMATION

**Corresponding Author**
**Jaroslav Vlček** - Department of Physics and NTIS-European Centre of Excellence, University of West Bohemia, Univerzitní 8, 30100 Plzeň, Czech Republic; orcid.org/0000-0003-2627-2074; Email: vlcek@kfy.zcu.cz

**Authors**
**Michal Kaufman** – Department of Physics and NTIS-European Centre of Excellence, University of West Bohemia, Univerzitní 8, 30100 Plzeň, Czech Republic; orcid.org/0009-0001-1733-2998
**Jiří Houška** – Department of Physics and NTIS-European Centre of Excellence, University of West Bohemia, Univerzitní 8, 30100 Plzeň, Czech Republic; orcid.org/0000-0002-4809-4128
**Sadoon Farrukh** – Department of Physics and NTIS-European Centre of Excellence, University of West Bohemia, Univerzitní 8, 30100 Plzeň, Czech Republic; orcid.org/0000-0002-3555-691X
**Stanislav Haviar** – Department of Physics and NTIS-European Centre of Excellence, University of West Bohemia, Univerzitní 8, 30100 Plzeň, Czech Republic; orcid.org/0000-0001-6926-8927

## ACKNOWLEDGMENTS

This work was supported by the Czech Science Foundation under Project No. 21-28277S and by the project Quantum materials for applications in sustainable technologies (QM4ST), funded as Project No. CZ.02.01.01/00/22_008/0004572 by Programme Johannes Amos Commenius, call Excellent Research.

**CONFLICT OF INTEREST**

The authors declare no conflict of interest.

**DATA AVAILABILITY STATEMENT**

The data that support the findings of this study are available from the corresponding author upon reasonable request.

TOC graphics

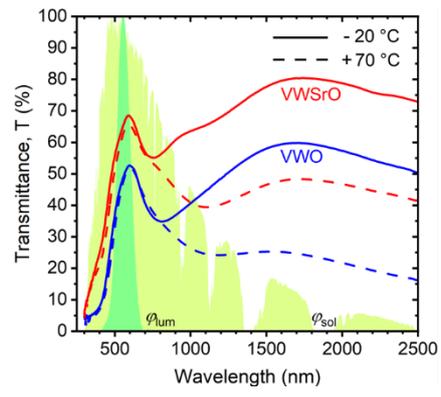